\documentclass{article}
\usepackage{amsfonts}
\usepackage{amsmath}

\input{tcilatex}
\begin{document}

\title{Path integral solution for a Klein-Gordon particle in vector and
scalar deformed radial Rosen-Morse-type potentials}
\author{A Khodja, A Kadja, F Benamira and L Guechi* \\
Laboratoire de Physique Th\'{e}orique, D\'{e}partement de Physique, \and %
Facult\'{e} des Sciences Exactes, Universit\'{e} des fr\`{e}res Mentouri,
\and Constantine 1, Route d'Ain El Bey, Constantine, Algeria}
\maketitle

\begin{abstract}
The problem of a Klein-Gordon particle moving in equal vector and scalar
Rosen-Morse-type potentials is solved in the framework of Feynman's path
integral approach. Explicit path integration leads to a closed form for the
radial Green's function associated with different shapes of the potentials.
For $q\leq -1$, and $\frac{1}{2\alpha }\ln \left\vert q\right\vert
<r<+\infty $, the energy equation and the corresponding wave functions are
deduced for the $l$ states using an appropriate approximation to the
centrifugal potential term. When $-1<q<0$ or $q>0$, it is shown that the
quantization conditions for the bound state energy levels $E_{n_{r}}$ are
transcendental equations which can be solved numerically. Three special
cases such as the standard radial Manning-Rosen potential $(\left\vert
q\right\vert =1)$, the standard radial Rosen-Morse potential $%
(V_{2}\rightarrow -V_{2},q=1)$ and the radial Eckart potential $%
(V_{1}\rightarrow -V_{1},q=1)$ are also briefly discussed.



*Corresponding author, E-mail: guechilarbi@yahoo.fr
\end{abstract}

\section{Introduction}

The purpose of the present paper is to further study the problem of a
spinless relativistic particle \ of mass $M$ and charge $(-e)$ in
interaction with a mixed central potential consisting of a vector potential $%
V_{q}\left( r\right) $ and a scalar potential $S_{q}\left( r\right) $. The
scalar potential is added to the rest mass and the whole can be interpreted
as an effective position dependent mass. The potentials $V_{q}\left(
r\right) $ and $S_{q}\left( r\right) $ are equal and of the form (see, for
example Ref. \cite{Kadja} and references therein)%
\begin{equation}
V_{q}\left( r\right) =S_{q}\left( r\right) =-\frac{V_{1}}{\cosh
_{q}^{2}(\alpha r)}-V_{2}\tanh _{q}(\alpha r),  \label{a.1}
\end{equation}%
where $V_{1}$, $V_{2}$, $\alpha $ and $q$ are four ajustable real
parameters, $V_{1}$ and $V_{2}$ describe the depth of the potential well, $%
\alpha $ is the screening parameter related to the range of the potential
characterized by the length $\alpha ^{-1}$. The expression (\ref{a.1}) is
defined in terms of $q-$deformed hyperbolic functions%
\begin{equation}
\sinh _{q}x=\frac{e^{x}-qe^{-x}}{2},\text{ \ \ \ \ }\cosh _{q}x=\frac{%
e^{x}+qe^{-x}}{2},\text{ \ \ \ \ }\tanh _{q}x=\frac{\sinh _{q}x}{\cosh _{q}x}%
\text{\ }  \label{a.1a}
\end{equation}%
which have been introduced for the first time by Arai \cite{Arai}. The
deformation parameter $q$ being a non-zero real number. The expression (\ref%
{a.1}) represents a variant of the deformed Schi\"{o}berg potential \cite%
{Amrouche} or the Tietz potential \cite{Zhang} which serve as modeling
potentials for diatomic molecules. In the case where $q=1$, the potentials (%
\ref{a.1}) reduce to the usual radial Rosen-Morse potential \ that has been
investigated from different points of view in the last decade \cite%
{Akoshile,Ikhdair1,Setare,Ibrahim,Aguda,Chen,Hamzavi,Tan,Liu}. Also, for $%
-1\leq q<0$ or $q>0$, various methods have been used to solve the
Klein-Gordon and Dirac equations \cite{Jia,Yang,Wang,Ikhdair2,Ghoumaid} with
these same potentials. In particular, for $\left\vert q\right\vert =1$ and
for a light modification of expression (\ref{a.1}), the relativistic
rotational-vibrational energies and the radial wave functions have been
approximately calculated with the help of the supersymmetric WKB approach
and through the resolution of the Klein-Gordon equation \cite{He}. More
recently, the usefulness of the Manning-Rosen potential and \ the
Rosen-Morse potential for calculations of the vibrational partition function
to study the thermodynamic properties of diatomic molecules has been
emphasized by Jia and co-workers \cite{Wang2,Song}.

In this study, we shall present a complete treatment of the bound state
problem for the potentials (\ref{a.1}) via the path integral approach for
any real value of parameter $q\neq 0$.

The organization of the paper is as follows: in section 2, we formulate the
radial Green's function associated with any $l-$wave in the framework of
Feynman's path integral. In section 3, we construct the radial Green's
function associated with the deformed Manning-Rosen potential $\left(
q<0\right) $. When $q\leq -1$ and $\frac{1}{\alpha }\ln \left\vert
q\right\vert <r<+\infty $, we use an appropriate approximation to the
centrifugal potential term to calculate the expression of the radial Green's
function for a state of orbital quantum number $l$. From this, we shall then
obtain the equation for the energy spectrum and the normalized
eigenfunctions. For $-1<q<0$, the $q-$deformed Manning-Rosen potential is
converted into the standard Manning-Rosen potential which is defined on a
half-line. In this case, the radial Green's function, for $l=0$, is
evaluated in a closed form by using the perturbation method which consists
in adding a Dirac $\delta -$function perturbation to the standard
Manning-Rosen potential and making the strength of this perturbation
infinitely repulsive to create a totally reflecting boundary. In this
fashion, we obtain the radial Green's function for a particle moving on a
half-line. From the poles of the Green's function, we derive a
transcendental equation for the energy levels. In section 4, the $q-$%
deformed Rosen-Morse potential $(q>0)$ is worked out in a similar way. We
first transform the path integral associated with this potential into the
one of the standard Rosen-Morse potential on a half-line and by means of the
same Dirac $\delta -$function perturbation trick, we calculate the radial
Green's function and also obtain a transcendental equation for the $s-$state
energy levels. In section 5, the standard radial Manning-Rosen potential $%
(\left\vert q\right\vert =1)$, the standard radial Rosen-Morse potential $%
(V_{2}\rightarrow -V_{2},$ $q=1)$ and the radial Eckart potential $%
(V_{1}\rightarrow -V_{1},$ $q=1)$ are considered as special cases. Section 6
will be a conclusion.

\section{Green's function}

The Green's function associated with a particle moving in a vector potential
and a scalar potential with spherical symmetry can develop into partial
waves \cite{Peak} in spherical coordinates:

\begin{equation}
G\left( \overset{\rightarrow }{r^{\prime \prime },t^{\prime \prime }},%
\overset{\rightarrow }{r^{\prime },t^{\prime }}\right) =\frac{1}{r^{\prime
\prime }\text{ }r^{\prime }\text{ }}\sum_{l=0}^{\infty }\frac{2l+1}{4\pi }%
G_{l}(r^{\prime \prime },t^{\prime \prime },r^{\prime },t^{\prime
})P_{l}\left( \cos \Theta \right) ,\,\,  \label{a.2}
\end{equation}%
where the radial Green's function is given by

\begin{eqnarray}
G_{l}(r^{\prime \prime },t^{\prime \prime },r^{\prime },t^{\prime }) &=&%
\frac{1}{2i}\dint\limits_{0}^{\infty }d\Lambda \text{ }\left\langle
r^{\prime \prime },t^{\prime \prime }\right\vert \exp \left[ \frac{i}{2}%
\left[ -P_{r}^{2}+(P_{0}-V_{q})^{2}\right. \right.  \notag \\
&&\left. \left. -\frac{l(l+1)}{r^{2}}-(M+S_{q})^{2}\right] \Lambda \right]
\left\vert r^{\prime },t^{\prime }\right\rangle ,  \label{a.3}
\end{eqnarray}%
and $P_{l}\left( \cos \Theta \right) $ is the Legendre polynomial of degree $%
l$ in $\cos \Theta =\frac{\overrightarrow{r}^{\prime \prime }.%
\overrightarrow{r}^{\prime }}{r^{\prime \prime }r^{\prime }}=\cos \theta
^{\prime \prime }\cos \theta ^{\prime }+\sin \theta ^{\prime \prime }\sin
\theta ^{\prime }\cos (\phi ^{\prime \prime }-\phi ^{\prime }).$

In Feynman's formulation \cite{Hibbs,Kleinert}, the radial Green's function $%
G_{l}(r^{\prime \prime },t^{\prime \prime },r^{\prime },t^{\prime })$ is
explicitly expressed in the form of a path integral,$\ $

\begin{equation}
G_{l}(r^{\prime \prime },t^{\prime \prime },r^{\prime },t^{\prime })=\frac{1%
}{2i}\dint\limits_{0}^{\infty }d\Lambda P_{l}(r^{\prime \prime },t^{\prime
\prime },r^{\prime },t^{\prime };\Lambda ),  \label{a.4}
\end{equation}%
where

\begin{eqnarray}
P_{l}(r^{\prime \prime },t^{\prime \prime },r^{\prime },t^{\prime };\Lambda
) &=&\text{ }\underset{N\rightarrow \infty }{\lim }\dprod\limits_{n=1}^{N}%
\left[ \dint dr_{n}dt_{n}\right] \text{ }\dprod\limits_{n=1}^{N+1}\left[
\dint \frac{d(P_{r})_{n}\text{ }d(P_{0})_{n}}{(2\pi )^{2}}\right]  \notag \\
&&\times \exp \left[ i\text{ }\dsum\limits_{n=1}^{N+1}A_{1}^{n}\right] ,
\label{a.5}
\end{eqnarray}%
with the short time action $A_{1}^{n}$ given by

\begin{eqnarray}
A_{1}^{n} &=&-(P_{r})_{n}\Delta r_{n}+(P_{0})_{n}\Delta t_{n}+\frac{%
\varepsilon _{\Lambda }}{2}\text{ }\left[ -(P_{r})_{n}^{2}+\left(
(P_{0})_{n}-V_{q}(r_{n})\right) ^{2}\right.  \notag \\
&&\left. -\frac{l(l+1)}{r_{n}^{2}}-(M+S_{q}(r_{n}))^{2}\right] ,  \label{a.6}
\end{eqnarray}%
in which the notation used is $\Delta r_{n}=r_{n}-r_{n-1},\Delta
t_{n}=t_{n}-t_{n-1},r_{n}=r(t_{n}),$ and $\varepsilon _{\Lambda }=\frac{%
\Lambda }{N+1}.$

Let us notice first that the integrations on the variables $t_{n}$ in the
expression (\ref{a.5}) give $N$ Dirac distributions $\delta \left(
(P_{0})_{n}-(P_{0})_{n+1}\right) $. Thereafter, performing the integrations
on the variables $(P_{0})_{n}$, one finds that

\begin{equation}
(P_{0})_{1}=(P_{0})_{2}=...=(P_{0})_{N+1}=E.  \label{a.7}
\end{equation}%
Consequently, the propagator (\ref{a.5}) becomes

\begin{equation}
P_{l}(r^{\prime \prime },t^{\prime \prime },r^{\prime },t^{\prime };\Lambda
)=\frac{1}{2\pi }\dint\limits_{-\infty }^{+\infty }dE\exp \left[
iE(t^{\prime \prime }-t^{\prime })\right] P_{l}(r^{\prime \prime },r^{\prime
};\Lambda ),  \label{a.8}
\end{equation}%
with the kernel $P_{l}(r^{\prime \prime },r^{\prime };\Lambda )$ given by

\begin{equation}
P_{l}(r^{\prime \prime },r^{\prime };\Lambda )=\text{ }\underset{%
N\rightarrow \infty }{\lim }\dprod\limits_{n=1}^{N}\left[ \dint dr_{n}\right]
\dprod\limits_{n=1}^{N+1}\left[ \dint \frac{d(P_{r})_{n}}{2\pi }\right] \exp %
\left[ i\dsum\limits_{n=1}^{N+1}A_{2}^{n}\right] ,  \label{a.9}
\end{equation}%
where the new short time action is then%
\begin{eqnarray}
A_{2}^{n} &=&-(P_{r})_{n}\Delta r_{n}+\frac{\varepsilon _{\Lambda }}{2}\text{
}\left[ -(P_{r})_{n}^{2}+(E-V_{q}(r_{n}))^{2}\right.  \notag \\
&&\left. -\frac{l(l+1)}{r_{n}^{2}}-(M+S_{q}(r_{n}))^{2}\right] .
\label{a.10}
\end{eqnarray}%
Then, performing the integrations over the variables $(P_{r})_{n},$ we find

\begin{eqnarray}
P_{l}(r^{\prime \prime },r^{\prime };\Lambda ) &=&\frac{1}{\sqrt{2i\pi
\varepsilon _{\Lambda }}}\text{ }\underset{N\rightarrow \infty }{\lim }%
\dprod\limits_{n=1}^{N}\left[ \dint \frac{dr_{n}}{\sqrt{2i\pi \varepsilon
_{\Lambda }}}\right] \exp \left\{ i\dsum\limits_{n=1}^{N+1}\left[ \frac{%
\left( \Delta r_{n}\right) ^{2}}{2\varepsilon _{\Lambda }}\right. \right. 
\notag \\
&&\left. \left. -\frac{\varepsilon _{\Lambda }}{2}\left( \left[
M+S_{q}(r_{n})\right] ^{2}-\left[ E-V_{q}(r_{n})\right] ^{2}+\frac{l(l+1)}{%
r_{n}^{2}}\right) \right] \right\} ,  \notag \\
&&  \label{a.11}
\end{eqnarray}%
and substituting (\ref{a.8}) into (\ref{a.4}), we notice that the term
depending on time $t$ does not contain the pseudo-time variable $\Lambda .$
Thus, we can rewrite the Green's function (\ref{a.4}) in the form: 
\begin{equation}
G_{l}(r^{\prime \prime },t^{\prime \prime },r^{\prime },t^{\prime })=\frac{1%
}{2\pi }\int_{-\infty }^{+\infty }dE\exp \left[ iE(t^{\prime \prime
}-t^{\prime })\right] G_{l}(r^{\prime \prime },r^{\prime }),  \label{a.12}
\end{equation}%
where

\begin{equation}
G_{l}(r^{\prime \prime },r^{\prime })=\frac{1}{2i}\int_{0}^{\infty }d\Lambda
P_{l}(r^{\prime \prime },r^{\prime };\Lambda ).  \label{a.13}
\end{equation}%
By assuming that $V_{q}(r)=S_{q}(r),$ the radial Green's function (\ref{a.13}%
) reduces to

\begin{equation}
G_{l}(r^{\prime \prime },r^{\prime })=\frac{1}{2i}\int_{0}^{\infty }d\Lambda
\exp \left( i\widetilde{E}_{0}^{2}\Lambda \right) K_{l}(r^{\prime \prime
},r^{\prime };\Lambda ),  \label{a.14}
\end{equation}%
where

\begin{eqnarray}
K_{l}(r^{\prime \prime },r^{\prime };\Lambda ) &=&\frac{1}{\sqrt{2i\pi
\varepsilon _{\Lambda }}}\underset{N\rightarrow \infty }{\lim }%
\dprod\limits_{n=1}^{N}\left[ \dint \frac{dr_{n}}{\sqrt{2i\pi \varepsilon
_{\Lambda }}}\right] \exp \left\{ i\dsum\limits_{n=1}^{N+1}\left[ \frac{%
\left( \Delta r_{n}\right) ^{2}}{2\varepsilon _{\Lambda }}\right. \right. 
\notag \\
&&\left. \left. -\frac{\varepsilon _{\Lambda }}{2}\left( 2\left( E+M\right)
V_{q}(r_{n})+\frac{l(l+1)}{r_{n}^{2}}\right) \right] \right\} ,  \label{a.15}
\end{eqnarray}%
and

\begin{equation}
\widetilde{E}_{0}^{2}=\frac{E^{2}-M^{2}}{2}.  \label{a.16}
\end{equation}%
The radial propagator (\ref{a.15}) and the radial Green's function (\ref%
{a.14}) depend on the arbitrary real deformation parameter $q$. When $-1<q<0$
or $q>0$, the radial Green's function (\ref{a.14}) can be only evaluated
exactly for the $s$ states, but, when $q\leq -1$, the $l-$state problem can
be solved by using a proper approximation to the centrifugal potential term.
So to undertake this study, three interesting cases must be distinguished
according to the values of the deformation parameter $q$.

\section{Deformed radial Manning-Rosen Potentials}

\subsection{First case: $q\leq -1.$}

When $q\leq -1$, the potentials (\ref{a.1}) are written in the form:

\begin{equation}
V_{q}\left( r\right) =S_{q}\left( r\right) =-\frac{V_{1}}{\sinh _{\left\vert
q\right\vert }^{2}(\alpha r)}-V_{2}\coth _{\left\vert q\right\vert }(\alpha
r).  \label{a.17}
\end{equation}

The motion takes place in the half-space $r>r_{0}=\frac{1}{2\alpha }\ln (-q)$%
. The figure 1 represents the variations with $(\alpha r)$ of the deformed
Manning-Rosen potential (\ref{a.17}) for three different $\left\vert
q\right\vert $ values. In order to construct the path integral for a state
of orbital quantum number $l$, we first use the expression 
\begin{equation}
\frac{1}{r^{2}}\approx \alpha ^{2}\left( \frac{1}{3}+\frac{\left\vert
q\right\vert }{\sinh _{\left\vert q\right\vert }^{2}(\alpha r)}\right)
\label{a.18}
\end{equation}%
as a good approximation for $\frac{1}{r^{2}}$ in the centrifugal potential
term when $\left\vert q\right\vert \geq 1$ as it can be seen in Fig. 2 which
contains a plot of $\frac{1}{(\alpha r)^{2}}$ and of $\left( \frac{1}{3}+%
\frac{\left\vert q\right\vert }{\sinh _{\left\vert q\right\vert }^{2}(\alpha
r)}\right) $ for some characteristic values of $\left\vert q\right\vert $.
Note that the approximation (\ref{a.18}) is equivalent to those of the
literature \cite{Cui,Guechi} for $\left\vert q\right\vert =1$. We next
transform the variable $r\in \left] r_{0},+\infty \right[ $ into a new
variable $x\in \left] 0,+\infty \right[ $ by

\begin{equation}
r=\frac{1}{\alpha }\left( x+\frac{1}{2}\ln \left\vert q\right\vert \right) .
\label{a.19}
\end{equation}%
In addition to the change of variable, we rescale the local time interval $%
\varepsilon _{\Lambda }$ to a new time interval $\alpha ^{-2}\varepsilon
_{s} $. Putting these considerations together, we can rewrite the Green's
function (\ref{a.14}) as: 
\begin{equation}
G_{l}(r^{\prime \prime },r^{\prime })=-\frac{1}{2\alpha }G_{MR}^{l}\left(
x^{\prime \prime },x^{\prime };\widetilde{E}_{l}^{2}\right) ,  \label{a.20}
\end{equation}%
with%
\begin{equation}
G_{MR}^{l}\left( x^{\prime \prime },x^{\prime };\widetilde{E}_{l}^{2}\right)
=i\int_{0}^{\infty }dS\exp \left( i\frac{\widetilde{E}_{l}^{2}}{\alpha ^{2}}%
S\right) K_{MR}^{l}\left( x^{\prime \prime },x^{\prime };S\right) ,
\label{a.21}
\end{equation}%
where%
\begin{equation}
\widetilde{E}_{l}^{2}=\widetilde{E}_{0}^{2}-\frac{l(l+1)\alpha ^{2}}{6},
\label{a.22a}
\end{equation}%
and%
\begin{equation}
K_{MR}^{l}\left( x^{\prime \prime },x^{\prime };S\right) =\int Dx(s)\exp
\left\{ i\int_{0}^{S}\left[ \frac{\overset{.}{x}^{2}}{2}-V_{MR}^{l}(x)\right]
ds\right\}  \label{a.22b}
\end{equation}%
is the propagator associated with the standard Manning-Rosen potential \cite%
{Manning}%
\begin{equation}
V_{MR}^{l}(x)=-\widetilde{V}_{2}\coth x+\frac{\widetilde{V}_{1}}{\sinh ^{2}x}%
;\text{ \ \ }x>0,  \label{a.23}
\end{equation}%
in which we have put%
\begin{equation}
\widetilde{V}_{1}=-(E+M)\frac{V_{1}}{\left\vert q\right\vert \alpha ^{2}}+%
\frac{l(l+1)}{2};\text{ \ \ \ \ \ \ \ }\widetilde{V}_{2}=(E+M)\frac{V_{2}}{%
\alpha ^{2}}.  \label{a.24}
\end{equation}

The Green's function can be stated in a closed form as is known from the
literature \cite{Grosche2,Benamira1,Benamira2}

\begin{eqnarray}
G_{MR}^{l}\left( x^{\prime \prime }\,\,,\,x^{\prime }\,;\widetilde{E\,}%
_{l}^{2}\,\right) &=&\frac{\Gamma (M_{1}-L_{E})\Gamma (L_{E}+M_{1}+1)}{%
\Gamma (M_{1}+M_{2}+1)\Gamma (M_{1}-M_{2}+1)}  \notag \\
&&\times \left( \frac{2}{1+\coth x^{\prime }}.\frac{2}{1+\coth x^{\prime
\prime }}\right) ^{\frac{M_{1}+M_{2}+1}{2}}  \notag \\
&&\times \text{ }\left( \frac{\coth x^{\prime }-1}{\coth x^{\prime }+1}.%
\frac{\coth x^{\prime \prime }-1}{\coth x^{\prime \prime }+1}\right) ^{\frac{%
M_{1}-M_{2}}{2}}  \notag \\
&&\times \text{ }_{2}F_{1}\left( M_{1}-L_{E},L_{E}+M_{1}+1,M_{1}-M_{2}+1;%
\frac{\coth x_{>}-1}{\coth x_{>}+1}\right)  \notag \\
&&\times \text{ }_{2}F_{1}\left( M_{1}-L_{E},L_{E}+M_{1}+1,M_{1}+M_{2}+1;%
\frac{2}{\coth x_{<}+1}\right) .  \notag \\
&&  \label{a.25}
\end{eqnarray}%
The quantities $L_{E},$ $M_{1}$ and $M_{2}$ are given by%
\begin{equation}
\left\{ 
\begin{array}{c}
L_{E}=-\frac{1}{2}+\frac{1}{2\alpha }\sqrt{(M+E)(M-E+2V_{2})+\frac{\alpha
^{2}}{3}l(l+1)}; \\ 
M_{1,2}=\sqrt{\left( l+\frac{1}{2}\right) ^{2}-2(M+E)\frac{V_{1}}{\alpha
^{2}\left\vert q\right\vert }}\pm \frac{1}{2\alpha }\sqrt{(M+E)(M-E-2V_{2})+%
\frac{\alpha ^{2}}{3}l(l+1)}.%
\end{array}%
\right.  \label{a.26}
\end{equation}

The energy spectrum for the bound states can be obtained from the poles of
the Green's function (\ref{a.25}) or the poles of the Euler function \ $%
\Gamma (M_{1}-L_{E})$. These poles are given by 
\begin{equation}
M_{1}-L_{E}=-n_{r};\text{ \ }n_{r}=0,1,2,...\text{ .}  \label{a.27}
\end{equation}%
By inserting the values of $L_{E}$ and $M_{1}$ in (\ref{a.27}), we obtain%
\begin{eqnarray}
M^{2}-E_{n_{r},l}^{2} &=&\frac{(M+E_{n_{r},l})^{2}V_{2}^{2}}{\alpha
^{2}\left( n_{r}+\frac{1}{2}+\sqrt{\left( l+\frac{1}{2}\right)
^{2}-2(M+E_{n_{r},l})\frac{V_{1}}{\alpha ^{2}\left\vert q\right\vert }}%
\right) ^{2}}  \notag \\
&&+\alpha ^{2}\left( n_{r}+\frac{1}{2}+\sqrt{\left( l+\frac{1}{2}\right)
^{2}-2(M+E_{n_{r},l})\frac{V_{1}}{\alpha ^{2}\left\vert q\right\vert }}%
\right) ^{2}  \notag \\
&&-\frac{\alpha ^{2}}{3}l(l+1).  \label{a.28}
\end{eqnarray}%
The corresponding wave functions are of the form:%
\begin{eqnarray}
u_{n_{r},l}^{q\leq -1}(r) &=&r\Psi _{n_{r},l}^{q\leq -1}(r)  \notag \\
&=&N_{n_{r},l}\left( 1-\left\vert q\right\vert e^{-2\alpha r}\right)
^{\delta _{l}}\left( \left\vert q\right\vert e^{-2\alpha r}\right)
^{w_{l}}P_{n_{r}}^{(2w_{l},2\delta _{l}-1)}\left( 1-2\left\vert q\right\vert
e^{-2\alpha r}\right) ,  \notag \\
&&  \label{a.29}
\end{eqnarray}%
where%
\begin{equation}
\left\{ 
\begin{array}{c}
w_{l}=\frac{1}{2\alpha }\sqrt{M^{2}-E_{n_{r},l}^{2}-2\left(
M+E_{n_{r},l}\right) V_{2}+\frac{\alpha ^{2}}{3}l(l+1)}; \\ 
\delta _{l}=\frac{1}{2}+\sqrt{\left( l+\frac{1}{2}\right)
^{2}-2(M+E_{n_{r},l})\frac{V_{1}}{\alpha ^{2}\left\vert q\right\vert }}.%
\end{array}%
\right.  \label{a.30}
\end{equation}%
The normalization constant $N_{n_{r}}$ follows from the condition%
\begin{equation}
\int_{\frac{1}{2\alpha }\ln \left\vert q\right\vert }^{+\infty }\left\vert
u_{n_{r},l}^{q\leq -1}(r)\right\vert ^{2}dr=1.  \label{a.31}
\end{equation}%
The calculation leads to%
\begin{equation}
N_{n_{r},l}=\left[ \frac{4\alpha w_{l}(n_{r}+w_{l}+\delta _{l})}{%
n_{r}+\delta _{l}}\frac{n_{r}!\Gamma (n_{r}+2w_{l}+2\delta _{l})}{\Gamma
(n_{r}+2w_{l}+1)\Gamma (n_{r}+2\delta _{l})}\right] ^{\frac{1}{2}}.
\label{a.32}
\end{equation}

\subsection{Second case: $-1<q<0$}

In this case, the potential (\ref{a.17}) is defined in the interval $%
\mathbb{R}
^{+}$. A plot of this potential for three different $\left\vert q\right\vert 
$ values is shown in Fig. 3. The transformation $r=\frac{1}{\alpha }(x+\frac{%
1}{2}\ln \left\vert q\right\vert )$ converts $r\in \left] 0,+\infty \right[ $
into $x\in \left] -\frac{1}{2}\ln \left\vert q\right\vert ,+\infty \right[ $%
. This means that the kernel (\ref{a.22b}) is the propagator describing the
evolution of a particle in the presence of a Manning-Rosen-type potential on
the half-line $x>x_{0}=$ $-\frac{1}{2}\ln \left\vert q\right\vert $. As we
can not perform a direct path integration to evaluate the propagator (\ref%
{a.22b}), the problem can be solved by a trick that consists in
incorporating an auxiliary term potential defined by a Dirac $\delta $
function in equation (\ref{a.22b}) to form an impenetrable barrier \cite%
{Clark} at $x=x_{0}$. Since the approximation (\ref{a.18}) is not suitable
for $0<\left\vert q\right\vert <1$, we limit ourselves to the evaluation of
the Green's function associated with the $s-$waves. So the Green's function (%
\ref{a.21}), for $l=0$, becomes 
\begin{equation}
G_{MR}^{\delta }\left( x^{\prime \prime },x^{\prime };\widetilde{E}%
_{0}^{2}\right) =i\int_{0}^{\infty }dS\exp \left( i\frac{\widetilde{E}%
_{0}^{2}}{\alpha ^{2}}S\right) K_{MR}^{\delta }\left( x^{\prime \prime
},x^{\prime };S\right) ,  \label{a.33}
\end{equation}%
where

\begin{equation}
K_{MR}^{\delta }\left( x^{\prime \prime },x^{\prime };S\right) =\int
Dx(s)\exp \left\{ i\int_{0}^{S}\left[ \frac{\overset{.}{x}^{2}}{2}%
-V_{MR}^{\delta }(x)\right] ds\right\} .  \label{a.34}
\end{equation}%
The path integral (\ref{a.34}) is the propagator of a particle which moves
in a potential of the form:

\begin{equation}
V_{MR}^{\delta }(x)=V_{MR}^{0}(x)-\eta \delta (x-x_{0}),  \label{a.35}
\end{equation}%
where $V_{MR}^{0}(x)$ is the expression of the potential (\ref{a.23}) for $%
l=0$. As is quite clear, given the complicated form of the potential (\ref%
{a.35}), the calculation of the Green's function (\ref{a.33}) can not be
performed directly. We propose to apply the perturbation approach in order
to express $\exp \left( i\eta \int_{s^{\prime }}^{s^{\prime \prime }}\delta
(x-x_{0})ds\right) $ in power series. Then, the propagator (\ref{a.34}) can
be written as:

\begin{eqnarray}
K^{\delta }\left( x^{\prime \prime }\,\,,\,x^{\prime }\,;\,S\,\right)
&=&K_{MR}^{0}\left( x^{\prime \prime }\,\,,\,x^{\prime }\,;\,S\,\right)
+\sum_{n=1}^{\infty }\frac{\left( i\eta \right) ^{n}}{n!}\overset{n}{%
\underset{j=1}{\Pi }}\left[ \dint_{s_{i}}^{s_{j+1}}ds_{j}\dint_{-\infty
}^{\infty }dx_{j}\right]  \notag \\
&&\times \text{ }K_{MR}^{0}\left( x_{1}\,\,,\,x^{\prime
}\,;\,s_{1}-s_{i}\,\right) \delta \left( x_{1}-x_{0}\right) \text{ }%
K_{MR}^{0}\left( x_{2}\,\,,x_{1}\,;\,s_{2}-s_{1}\,\right)  \notag \\
&&\times ...\times \delta \left( x_{n-1}-x_{0}\right) \text{ }%
K_{MR}^{0}\left( x_{n}\,\,,x_{n-1}\,;\,s_{n}-s_{n-1}\,\right)  \notag \\
&&\times \delta \left( x_{n}-x_{0}\,\right) \text{ }K_{MR}^{0}\left(
x^{\prime \prime }\,\,,\,x_{n}\,;\,S-s_{n}\,\right)  \notag \\
&=&K_{MR}^{0}\left( x^{\prime \prime }\,\,,x^{\prime }\,;\,S\,\right)
+\sum_{n=1}^{\infty }\left( i\eta \right)
^{n}\dint_{s_{i}}^{s_{f}}ds_{n}\dint_{s_{i}}^{s_{n}}ds_{n-1}...%
\dint_{s_{i}}^{s_{2}}ds_{1}  \notag \\
&&\times \text{ }K_{MR}^{0}\left( x_{0}\,\,,x^{\prime
}\,;\,s_{1}-s_{i}\,\right) K_{MR}^{0}\left(
x_{0}\,\,,\,x_{0}\,;\,s_{2}-s_{1}\,\right) \times ...  \notag \\
&&\times \text{ }K_{MR}^{0}\left(
x_{0}\,\,,\,x_{0}\,;\,s_{n}-s_{n-1}\,\right) K_{MR}^{0}\left( x^{\prime
\prime }\,,x_{0}\,;S-s_{n}\,\,\right) ,  \label{a.36}
\end{eqnarray}%
where we ordered the time as follows: $s^{\prime
}=s_{0}<s_{1}<s_{2}<...<s_{n}<s_{n+1}=s^{\prime \prime }$. To perform the
successive integrations on the variables $s_{j}$ in equation (\ref{a.36}),
we insert (\ref{a.36}) into (\ref{a.33}) and use the convolution theorem of
Fourier transformation, then we obtain

\begin{equation}
G_{MR}^{\delta }\left( x^{\prime \prime }\,,\,x^{\prime }\,;\,\tilde{E}%
\,_{0}^{2}\right) =G_{MR}^{0}\left( x^{\prime \prime }\,,\,x^{\prime }\,;\,%
\tilde{E}_{0}^{2}\,\right) -\frac{G_{MR}^{0}\left( x^{\prime \prime
}\,,\,x_{0}\,;\,\tilde{E}_{0}^{2}\,\right) G_{MR}^{0}\left(
x_{0}\,,\,x^{\prime }\,;\,\tilde{E}_{0}^{2}\,\right) }{G_{MR}^{0}\left(
x_{0}\,,\,x_{0}\,;\,\tilde{E}_{0}^{2}\,\right) -\frac{1}{\eta }},
\label{a.37}
\end{equation}%
where $G_{MR}^{0}\left( x^{\prime \prime }\,,\,x^{\prime }\,;\,\tilde{E}%
_{0}^{2}\,\right) $ is the Green's function (\ref{a.25}) associated with the
standard Manning-Rosen potential (\ref{a.23}), for $l=0$.

If we now take the limit $\eta \rightarrow -\infty $, the physical system
will be forced to move in the potential $V_{MR}^{0}\left( x\right) $ bounded
by an infinitely repulsive barrier \cite{Grosche2,Clark} located at $%
x\,\,=\,x_{0}$. In this case, the Green's function is then given by:

\begin{eqnarray}
\tilde{G}_{MR}^{0}\left( x^{\prime \prime }\,\,,x^{\prime }\,;\,\tilde{E}%
\,_{0}^{2}\right) &=&\lim_{\eta \rightarrow -\infty }G_{MR}^{\delta }\left(
x^{\prime \prime }\,\,,\,x^{\prime }\,;\,\tilde{E}\,_{0}^{2}\right)  \notag
\\
&=&G_{MR}^{0}\left( x^{\prime \prime }\,\,,\,x^{\prime }\,;\,\tilde{E}%
_{0}^{2}\,\right) -\frac{G_{MR}^{0}\left( x^{\prime \prime }\,\,,x_{0}\,;\,%
\tilde{E}_{0}^{2}\,\right) G_{MR}^{0}\left( x_{0}\,\,,x^{\prime }\,;\,\tilde{%
E}_{0}^{2}\,\right) }{G_{MR}^{0}\left( x_{0}\,\,,\,x_{0}\,;\,\tilde{E}%
_{0}^{2}\,\right) }\text{ }.  \notag \\
&&  \label{a.38}
\end{eqnarray}%
Finally, when $-1<q<0$, the radial Green's function of our problem is
expressed as:%
\begin{equation}
G_{0}(r^{\prime \prime },r^{\prime })=-\frac{1}{2\alpha }\tilde{G}%
_{MR}^{0}\left( x^{\prime \prime }\,\,,x^{\prime }\,;\,\tilde{E}%
\,_{0}^{2}\right) .  \label{a.39}
\end{equation}

The energy spectrum is determined from the poles of the expression (\ref%
{a.38}), i.e., by the equation $G_{MR}^{0}\left( x_{0}\,\,,\,x_{0}\,;\,%
\tilde{E}_{0}^{2}\,\right) =0$, or as well by the transcendental equation%
\begin{equation}
_{2}F_{1}(\delta +w-p,\delta +p+w,2w+1;\left\vert q\right\vert )=0,
\label{a.40}
\end{equation}%
where the quantities $\delta $, $p$ and $w$ are defined as%
\begin{equation}
\left\{ 
\begin{array}{c}
\delta =\frac{1}{2}\left( 1+\sqrt{1-\frac{8(E_{n_{r}}+M)V_{1}}{\alpha
^{2}\left\vert q\right\vert }}\right) ; \\ 
p=\frac{1}{2\alpha }\sqrt{(M+E_{n_{r}})(M-E_{n_{r}}+2V_{2})}; \\ 
w=\frac{1}{2\alpha }\sqrt{(M+E_{n_{r}})(M-E_{n_{r}}-2V_{2})}.%
\end{array}%
\right.  \label{a.41}
\end{equation}%
The equation (\ref{a.40}) can be solved numerically to determine the
discrete energy levels of the particle. The corresponding wave functions are
of the form:%
\begin{eqnarray}
u_{n_{r}}^{-1<q<0}(r) &=&r\Psi _{n_{r}}^{-1<q<0}(r)=C\left( 1-\left\vert
q\right\vert e^{-2\alpha r}\right) ^{\delta }\left( \left\vert q\right\vert
e^{-2\alpha r}\right) ^{w}  \notag \\
&&\times \text{ }_{2}F_{1}(\delta +w-p,\delta +p+w,2w+1;\left\vert
q\right\vert e^{-2\alpha r}),  \label{a.42}
\end{eqnarray}%
where $C$ is a constant factor. Note that these wave functions well satisfy
the boundary conditions%
\begin{equation}
u_{n_{r}}^{-1<q<0}(r)\underset{\text{ }r\rightarrow 0}{\rightarrow }0\text{,
\ \ }  \label{a.43}
\end{equation}%
and%
\begin{equation}
u_{n_{r}}^{-1<q<0}(r)\underset{r\rightarrow \infty }{\rightarrow }0\text{.}
\label{a.44}
\end{equation}

\section{Deformed radial Rosen-Morse potentials}

For $q>0$, the potentials (\ref{a.1}) have the form of the deformed
Rosen-Morse potential which is defined in the interval $%
\mathbb{R}
^{+}$. Figure 4 contains a plot of the deformed radial Rosen-Morse potential
for six different $q$ values. In order to bring back the integral (\ref{a.14}%
), for $l=0$, to a solvable form, we proceed as in the previous case. We
perform the following coordinate transformation:

\begin{equation}
r\in 
\mathbb{R}
^{+}\rightarrow x\in \left] -\frac{1}{2}\ln q,+\infty \right[  \label{a.45}
\end{equation}

\bigskip defined by%
\begin{equation}
r=\frac{1}{\alpha }\left( x+\frac{1}{2}\ln q\right) .  \label{a.46}
\end{equation}

\bigskip After changing $\varepsilon _{\Lambda }$ into $\alpha
^{-2}\varepsilon _{s}$ or $\Lambda $ into $\alpha ^{-2}S$, we can write (\ref%
{a.14}), for the $s$ states, in the following form:%
\begin{equation}
G_{0}(r^{\prime \prime },r^{\prime })=-\frac{1}{2\alpha }\tilde{G}%
_{RM}^{0}\left( x^{\prime \prime }\,\,,x^{\prime }\,;\,\tilde{E}%
\,_{0}^{2}\right) ,  \label{a.47}
\end{equation}%
where 
\begin{equation}
\tilde{G}_{RM}^{0}\left( x^{\prime \prime }\,\,,x^{\prime }\,;\,\tilde{E}%
\,_{0}^{2}\right) =i\int_{0}^{\infty }dS\exp \left( i\frac{\widetilde{E}%
_{0}^{2}}{\alpha ^{2}}S\right) K_{RM}^{0}\left( x^{\prime \prime },x^{\prime
};S\right) ,  \label{a.48}
\end{equation}%
and%
\begin{equation}
K_{RM}^{0}\left( x^{\prime \prime },x^{\prime };S\right) =\int Dx(s)\exp
\left\{ i\int_{0}^{S}\left[ \frac{\overset{.}{x}^{2}}{2}-\widetilde{V}%
_{2}\tanh x+\frac{\widetilde{V}_{1}}{q\cosh ^{2}x}\right] ds\right\} .
\label{a.49}
\end{equation}%
The constants $\widetilde{V}_{1}$ and $\widetilde{V}_{2}$ are given by%
\begin{equation}
\widetilde{V}_{1}=(E+M)\frac{V_{1}}{\alpha ^{2}};\text{ \ \ \ \ \ \ \ }%
\widetilde{V}_{2}=-(E+M)\frac{V_{2}}{\alpha ^{2}}.  \label{a.50}
\end{equation}%
The propagator (\ref{a.49}) has the same shape as the path integral relative
to the potential originally introduced by Rosen and Morse to discuss the
vibrational states of the polyatomic molecules \cite{Rosen} . The
Rosen-Morse potential is defined for $x\in 
\mathbb{R}
$, but in this case we have transformed the path integral for the potentials
(\ref{a.1}) into a path integral for a standard Rosen-Morse-type potential
via the transformation $r\rightarrow r(x)$ which converts $r\in 
\mathbb{R}
^{+}\rightarrow x\in ]-(1/2)\ln q,+\infty \lbrack $. This means that the
motion of the particle takes place in the half-space $x>x_{0}=-(1/2)\ln q$.
Then, to calculate the Green's function relative to the $s-$waves, we
proceed as in the previous case and we obtain

\begin{eqnarray}
\tilde{G}_{RM}^{0}\left( x^{\prime \prime }\,\,,x^{\prime }\,;\,\tilde{E}%
\,_{0}^{2}\right) &=&G_{RM}^{0}\left( x^{\prime \prime }\,\,,\,x^{\prime
}\,;\,\tilde{E}_{0}^{2}\,\right) -\frac{G_{RM}^{0}\left( x^{\prime \prime
}\,\,,x_{0}\,;\,\tilde{E}_{0}^{2}\,\right) G_{RM}^{0}\left(
x_{0}\,\,,\,x^{\prime }\,;\,\tilde{E}_{0}^{2}\,\right) }{G_{RM}^{0}\left(
x_{0}\,\,,\,x_{0}\,;\,\tilde{E}_{0}^{2}\,\right) },\text{ }  \notag \\
&&  \label{a.51}
\end{eqnarray}%
where $G_{RM}^{0}\left( x^{\prime \prime }\,\,,\,x^{\prime }\,;\,\tilde{E}%
_{0}^{2}\,\right) $ is the Green's function associated with the standard
Rosen-Morse potential \cite{Rosen}%
\begin{equation}
V_{RM}(x)=\widetilde{V}_{2}\tanh x-\frac{\widetilde{V}_{1}}{q\cosh ^{2}x};%
\text{ \ \ \ \ }x\in 
\mathbb{R}
.  \label{a.52}
\end{equation}%
It is known that the solution by the path integral for this potential leads
to the following expression of the Green's function \cite%
{Grosche2,Benamira1,Benamira2}

\begin{eqnarray}
G_{RM}^{0}\left( x^{\prime \prime }\,\,,\,x^{\prime }\,;\,\tilde{E}%
_{0}^{2}\,\right) &=&\frac{\Gamma (M_{1}-L_{E})\Gamma (L_{E}+M_{1}+1)}{%
\Gamma (M_{1}+M_{2}+1)\Gamma (M_{1}-M_{2}+1)}  \notag \\
&&\times \text{ \ }\left( \frac{1-\tanh x^{\prime }}{2}\frac{1-\tanh
x^{\prime \prime }}{2}\right) ^{\left( M_{1}+M_{2}\right) /2}  \notag \\
&&\times \text{ \ }\left( \frac{1+\tanh x^{\prime }}{2}\frac{1+\tanh
x^{\prime \prime }}{2}\right) ^{\left( M_{1}-M_{2}\right) /2}  \notag \\
&&\times \text{ \ }_{\text{ }2}F_{1}\text{ }\left( M_{1}-L_{E},\text{ }%
L_{E}+M_{1}+1,\text{ }M_{1}-M_{2}+1;\text{ }\frac{1+\tanh x_{>}}{2}\right) 
\notag \\
&&\times \text{ \ }_{\text{ }2}F_{1}\text{ }\left( M_{1}-L_{E},\text{ }%
L_{E}+M_{1}+1,\text{ }M_{1}+M_{2}+1;\text{ }\frac{1-\tanh x_{<}}{2}\right) ,
\notag \\
&&  \label{a.53}
\end{eqnarray}%
with the notation%
\begin{equation}
\left\{ 
\begin{array}{c}
L_{E}=-\frac{1}{2}+\frac{1}{2}\sqrt{1+8\frac{(E+M)V_{1}}{\alpha ^{2}q}}; \\ 
M_{1,2}=\frac{1}{2\alpha }\left( \sqrt{(M+E)(M-E+2V_{2})}\pm \sqrt{%
(M+E)(M-E-2V_{2})}\right) .%
\end{array}%
\right.  \label{a.54}
\end{equation}

The bound state energy levels are determined from the poles of the Green's
function (\ref{a.51}), i.e., by the equation $G_{RM}^{0}\left(
x_{0}\,\,,\,x_{0}\,;\,\tilde{E}_{0}^{2}\,\right) =0,$ or as well by the
following quantization condition which is a transcendental equation
involving the hypergeometric function

\begin{equation}
_{2}F_{1}\left( p+w-\delta +1,p+w+\delta ,2p+1;\frac{1}{1+q}\right) =0,
\label{a.55}
\end{equation}%
where the parameters $\delta $, $p$ and $w$ are defined by%
\begin{equation}
\left\{ 
\begin{array}{c}
\delta =\frac{1}{2}\left( 1+\sqrt{1+8\frac{(E_{n_{r}}+M)V_{1}}{\alpha ^{2}q}}%
\right) ; \\ 
p=\frac{1}{2\alpha }\sqrt{(M+E_{n_{r}})(M-E_{n_{r}}-2V_{2})}; \\ 
w=\frac{1}{2\alpha }\sqrt{(M+E_{n_{r}})(M-E_{n_{r}}+2V_{2})}.%
\end{array}%
\right.  \label{a.56}
\end{equation}%
The equation (\ref{a.55}) can be also solved numerically.

Using the Green's function (\ref{a.53}) for the Rosen-Morse potential and
the link between (\ref{a.47}) and (\ref{a.53}), we show that the wave
functions corresponding to the bound states have the form:

\begin{eqnarray}
u_{n_{r}}^{q>0}(r) &=&r\Psi _{n_{r}}^{q>0}(r)=C\left( \frac{1}{%
1+qe^{-2\alpha r}}\right) ^{p}\left( \frac{q}{q+e^{2\alpha r}}\right) ^{w} 
\notag \\
&&\times \text{ }_{2}F_{1}\left( p+w-\delta +1,p+w+\delta ,2p+1;\frac{1}{%
1+qe^{-2\alpha r}}\right) ,  \label{a.57}
\end{eqnarray}%
where $C$ is a constant factor. These wave functions are physically
acceptable since they satisfy the boundary conditions 
\begin{equation}
u_{n_{r}}^{q>0}(r)\underset{\text{ }r\rightarrow 0}{\rightarrow }0,
\label{a.58}
\end{equation}%
and

\begin{equation}
u_{n_{r}}^{q>0}(r)\underset{\text{ }r\rightarrow \infty }{\rightarrow }0.
\label{a.59}
\end{equation}

\section{Particular cases}

\subsection{First case: standard radial Manning-Rosen potentials}

If we take $\left\vert q\right\vert =1$, the potentials (\ref{a.17}) turn to
the standard radial Manning-Rosen potential

\begin{equation}
V\left( r\right) =S\left( r\right) =-\frac{V_{1}}{\sinh ^{2}(\alpha r)}%
-V_{2}\coth (\alpha r).  \label{a.60}
\end{equation}%
The energy equation and the normalized wave functions of the bound states
can be deduced from expressions (\ref{a.28}) and (\ref{a.29})

\begin{eqnarray}
M^{2}-E_{n_{r},l}^{2} &=&\frac{(M+E_{n_{r},l})^{2}V_{2}^{2}}{\alpha
^{2}\left( n_{r}+\frac{1}{2}+\sqrt{\left( l+\frac{1}{2}\right)
^{2}-2(M+E_{n_{r},l})\frac{V_{1}}{\alpha ^{2}}}\right) ^{2}}  \notag \\
&&+\alpha ^{2}\left( n_{r}+\frac{1}{2}+\sqrt{\left( l+\frac{1}{2}\right)
^{2}-2(M+E_{n_{r},l})\frac{V_{1}}{\alpha ^{2}}}\right) ^{2}  \notag \\
&&-\frac{\alpha ^{2}}{3}l(l+1),  \label{a.61}
\end{eqnarray}

\begin{eqnarray}
u_{n_{r},l}^{\left\vert q\right\vert =1}(r) &=&r\Psi _{n_{r},l}^{\left\vert
q\right\vert =1}(r)  \notag \\
&=&\widetilde{N}_{n_{r},l}\left( 1-e^{-2\alpha r}\right) ^{\widetilde{\delta 
}_{l}}\left( e^{-2\alpha r}\right) ^{w_{l}}P_{n_{r}}^{(2w_{l},2\widetilde{%
\delta }_{l}-1)}\left( 1-2e^{-2\alpha r}\right) ,  \notag \\
&&  \label{a.62}
\end{eqnarray}%
where

\begin{equation}
\left\{ 
\begin{array}{c}
w_{l}=\frac{1}{2\alpha }\sqrt{M^{2}-E_{n_{r},l}^{2}-2\left(
M+E_{n_{r},l}\right) V_{2}+\frac{\alpha ^{2}}{3}l(l+1)}; \\ 
\widetilde{\delta }_{l}=\frac{1}{2}+\sqrt{\left( l+\frac{1}{2}\right)
^{2}-2(M+E_{n_{r},l})\frac{V_{1}}{\alpha ^{2}}}.%
\end{array}%
\right.  \label{a.63}
\end{equation}%
and

\begin{equation}
\widetilde{N}_{n_{r},l}=\left[ \frac{4\alpha w_{l}(n_{r}+w_{l}+\widetilde{%
\delta }_{l})}{n_{r}+\widetilde{\delta }_{l}}\frac{n_{r}!\Gamma
(n_{r}+2w_{l}+2\widetilde{\delta }_{l})}{\Gamma (n_{r}+2w_{l}+1)\Gamma
(n_{r}+2\widetilde{\delta }_{l})}\right] ^{\frac{1}{2}}.  \label{a.64}
\end{equation}

\subsection{Second case: standard radial Rosen-Morse potentials}

By setting $q=1$, and by changing $V_{2}$ in $\left( -V_{2}\right) ,$ the
expression (\ref{a.1}) becomes the so-called standard Rosen-Morse potential

\begin{equation}
V\left( r\right) =S\left( r\right) =-\frac{V_{1}}{\cosh ^{2}(\alpha r)}%
+V_{2}\tanh (\alpha r).  \label{a.65}
\end{equation}%
The energy levels $E_{n_{r}}$ are deduced from (\ref{a.55})\ by the
transcendental equation

\begin{equation}
_{2}F_{1}\left( p+w-\delta +1,p+w+\delta ,2p+1;\frac{1}{2}\right) =0,
\label{a.66}
\end{equation}%
and the non-normalized wave functions (\ref{a.57}) become in this case:

\begin{eqnarray}
u_{n_{r}}^{q=1}(r) &=&r\Psi _{n_{r}}^{q=1}(r)=C\left( \frac{1}{1+e^{-2\alpha
r}}\right) ^{p}\left( \frac{1}{1+e^{2\alpha r}}\right) ^{w}  \notag \\
&&\times \text{ }_{2}F_{1}\left( p+w-\delta +1,p+w+\delta ,2p+1;\frac{1}{%
1+e^{-2\alpha r}}\right) ,  \label{a.67}
\end{eqnarray}%
where $p$ is replaced by $w$ and conversely.

\subsection{Third case: radial Eckart potentials}

By taking $q=1$, and by changing $V_{1}$ in $\left( -V_{1}\right) $, the
potentials (\ref{a.1}) reduce to the Eckart potential

\begin{equation}
V\left( r\right) =S\left( r\right) =\frac{V_{1}}{\cosh ^{2}(\alpha r)}%
-V_{2}\tanh (\alpha r).  \label{a.68}
\end{equation}

The quantization condition of energy levels and the non-normalized wave
functions can be derived from equations (\ref{a.55}) and (\ref{a.57}). They
are written respectively,

\begin{equation}
_{2}F_{1}\left( p+w-\overline{\delta }+1,p+w+\overline{\delta },2p+1;\frac{1%
}{2}\right) =0,  \label{a.69}
\end{equation}%
and

\begin{eqnarray}
u_{n_{r}}^{q=1}(r) &=&r\Psi _{n_{r}}^{q=1}(r)=C\left( \frac{1}{1+e^{-2\alpha
r}}\right) ^{p}\left( \frac{1}{1+e^{2\alpha r}}\right) ^{w}  \notag \\
&&\times \text{ }_{2}F_{1}\left( p+w-\overline{\delta }+1,p+w+\overline{%
\delta },2p+1;\frac{1}{1+e^{-2\alpha r}}\right) ,  \label{a.70}
\end{eqnarray}%
with $\overline{\delta }=\frac{1}{2}+\sqrt{1-8\frac{(E_{n_{r}}+M)V_{1}}{%
\alpha ^{2}}}.$

\section{Conclusion}

In this article, we have solved the problem of a relativistic spinless
particle in the presence of equal vector and scalar $q-$deformed radial
Rosen-Morse-type potentials by path integration. This problem has only
partially been discussed through the resolution of the Klein-Gordon equation 
\cite{Jia}. However, a complete solution can be given \ for any deformation
parameter $q\neq 0$. As we have shown, the path integral for the Green's
function associated with this mixture of \ equal vector and scalar
potentials can not be evaluated in a unified manner whatever the value of
the deformation parameter. When $q\leq -1$ and $\frac{1}{\alpha }\ln
\left\vert q\right\vert <r<\infty $, the radial Green's function for any $l$
state is directly calculated by using an appropriate approximation to the
centrifugal potential term. The energy equation and the corresponding wave
functions are then obtained. For $-1<q<0$ or $q>0$, we have limited
ourselves to the evaluation of the Green's functions for the $s$ waves $%
(l=0) $. We have shown that the transformation of the Green's function
relative to the starting potential defined on the interval $%
\mathbb{R}
^{+}$ into a path integral associated with a Manning-Rosen or
Rosen--Morse-type potential reduces the problem to that of a particle forced
to move in a half-space $x>x_{0}$. This problem with the Dirichlet boundary
conditions is treated by using the perturbation approach. In both cases, the
quantization conditions are transcendental equations involving the
hypergeometric function which can be solved numerically to determine the
energy levels of the bound states. The radial wave functions, expressed in
terms of the hypergeometric functions are also derived.

\end{document}